\documentclass[]{article}

\def\be{\begin{equation}}
\def\te{\end{equation}}
\def\ee{\end{equation}}
\def\ba{\begin{eqnarray}}
\def\bea{\begin{eqnarray}}

\def\tea{\end{eqnarray}}
\def\ea{\end{eqnarray}}
\def\eea{\end{eqnarray}}

\newskip\humongous \humongous=0pt plus 1000pt minus 1000pt

\newif\ifdtup

\begin{document}

\title {Decoherence in Quantum Gravity: \\ Issues and Critiques}
\author{C.~Anastopoulos \footnote{Email address:
anastop@physics.upatras.gr}, \\ Department of Physics, University of
Patras, \\ 26500 Patras, Greece, \\ \\
  and B.~L. Hu\footnote{Email
 address: blhu@umd.edu}, \\ Department of
Physics, University of Maryland, \\ College Park, Maryland
20742-4111 }

\date{\today}
\maketitle

\begin{abstract}
An increasing number of papers have appeared in recent years on
decoherence in quantum gravity at the Planck energy.  We discuss the
meaning of decoherence in quantum gravity starting from the common
notion that quantum gravity is a theory for the microscopic
structures of spacetime, and invoking some generic features of
quantum decoherence from the open systems viewpoint.  We dwell on a
range of issues bearing on this process including the relation
between statistical and quantum, noise from effective field theory,
the meaning of stochasticity, the origin of non-unitarity and the
nature of nonlocality in this and related contexts. To expound these
issues we critique on two representative theories: One
\cite{Mavr,Mavr2} claims that decoherence in quantum gravity scale
leads to the violation of CPT symmetry at sub-Planckian energy which
is used to explain today's particle phenomenology. The other
\cite{Garay1,Garay2} uses this process in place with the Brownian
motion model to prove that spacetime foam behaves like a thermal
bath. A companion paper \cite{IntDecAH} will deal with intrinsic and
fundamental decoherence which also bear on issues in classical and
quantum gravity.
\end{abstract}


\newpage

\section{Introduction}

\subsection{Symnopsis of BLH's talk at this conference}

The title of the invited talk delivered by one of us (BLH) in this
meeting which this article stems from, was, "What is Quantum
Gravity?" (QG). It presents the view that if one agrees that the
common goal of QG is to search for a microscopic structure of
spacetime, then perhaps one should place more emphasis on the
macroscopic (M) to microscopic (m) issues than the classical (C) to
quantum (Q) issues.  The classical to quantum route  (C $\rightarrow$
Q), such as finding a good set of variables in general relativity and
quantizing them, is the way traditional quantum gravity effort has
been placed, epitomizing in the current loop QG program. (See
\cite{AppQG} for a collection of reviews on the various approaches to
quantum gravity.) We advocate the importance, or even precedence, of
the M $\rightarrow$ m route over the C $\rightarrow$ Q route: We need
to find out the appropriate microscopic constituents before
quantizing them. Or, in this viewpoint, one is doubtful that
quantizing the macroscopic variables in a theory valid perhaps only
at low energies, {\it vis}, the theory of general relativity (GR) can
lead to a theory of microscopic structure of spacetime.

This view is shared with string theory; the difference is that, being
less ambitious or brave (the `half full' view is being more prudent
and cautious), we take a `bottom-up' rather than a `top-down'
approach. We build our intuitions on and seek our way from reliable
physics at low energies, with the well established theories of
semiclassical gravity and the newer stochastic gravity \cite{stogra}
(an extension of quantum field theory in a curved classical
background \cite{BirDav} to include the expectation values of the
stress energy tensor of quantum matter fields  and their fluctuations
as sources). In this approach one looks from the given macroscopic
spacetime interacting with matter for the high energy relics (from
experiments or observations) or suggestive hints (from theoretical
constructs or conceptual reasons) of such microscopic structures.
Admittedly this is a very difficult if not impossible task, but this
is no different from most experimental and theoretical explorations
in the history of physics.

Going from the micro to the macro and from the few to the many, we
rely more on the tools and concepts of nonequilibrium statistical
mechanics including critical phenomena, hydrodynamics and kinetic
theory, aided by analog models and ideas from condensed matter
physics. The two keys the speaker thinks are of most use (for us low
energy creatures) to decipher the mysteries of the microscopic
structures at  higher energy scales are noise and topology, because
colored noise (correlation-fluctuations) ingrains the nature of the
environment and its influence on the low energy system of interest,
and topological structures can better survive the re-constituting
(including physical changes such as phase transition and descriptive
adjustments such as the adoption of better collective variables) and
unavoidably corruptive processes (in the sense of degradation of
information by coarse-graining as one traverses different layers of
structure and levels of interactions) in the time evolution (or
energy scaling) of the system over history.

One earlier proposal of the speaker is to view GR as hydrodynamics of
QG \cite{GRhydro} (regarded as a theory for the microscopic
structures of spacetime). So, given a hydrodynamic theory our task
would be to deduce the molecular dynamics and from it decipher the
properties of the molecules. It is easier to go from the micro to the
macro but there are ways to get some hints in the reverse direction,
such as examining the hydrodynamic fluctuations and nonperturbative
structures, and for these, critical phenomena, kinetic theory and
stochastic processes can be of help.  We want to know how and where
different layers of structure and their interactions can emerge in
the interceding processes.   In this quest two conceptual pathways
have been explored, the kinetic theory approach \cite{kinQG} and
spacetime as a condensate \cite{STcond}. One can find a summary of
these ideas in a recent essay of the speaker \cite{nvQG-OU} which
also uses this view to address the so-called `origin' of the universe
issue \footnote{It also contains a short bibliography of some latest
papers in related approaches to QG. Amongst the different approaches
to quantum gravity  the issues expounded by Sorkin \cite{Sorkin}, the
ideas proposed by Wen \cite{Wen} based on quantum order, the analog
to condensed matter systems as expounded by Volovik \cite{Volovik},
and the programs pursued by Ambjorn and Loll on Lorenzian dynamics of
triangulated spacetime \cite{Loll}, that of Dreyer, Friedel, Levine,
Markopoulo, Oriti, Rovelli and Smolin \cite{Rovelli,spinet} on the
structure and evolution of spin network are of particular interest,
because one can use these explicit constructions to examine the
issues raised here, e.g., seeing the hydrodynamic limit, or even the
dynamically preferred dimension-four spacetimes. Read the articles by
these authors collected in Oriti's book \cite{AppQG}. See also work
by Herzog, Son et al on the hydrodynamic limit of string theory and
its features \cite{Herzog,Son}.}.

\subsection{Aim of this work}

Instead of repeating what was said and written before on this topic
we would like to pick one representative issue in quantum gravity
which bears on several aspects touched on in the talk, specifically,
here, on decoherence in quantum gravity. This is an issue which has
drawn increasing recent attention from researchers interested in the
foundation of both quantum mechanics (such as intrinsic decoherence)
-- see e. g., Adler's book \cite{Adler} and its bibliography, and
general relativity (such as gravitational decoherence -- see e.g.,
\cite{KokYut}), and those who are interested in exotic yet poorly
understood (or even ill-defined) constructs such as spacetime or
quantum foams, or by conjectured universal mechanisms, such as
decoherence in quantum gravity. Usually one needs to draw on some
features out of the familiar, like nonunitarity and nonlocality, to
make new proposals work.  One aim of this paper is to emphasize that
one needs to check proposals against some known facts from detailed
model studies of specific issues (e.g, quantum decoherence
\cite{qdec}) or well-established mathematical (e.g., probability
theory) or physical theorems. By clarifying the underlying ideas in
such proposals, we hope to get a clearer view to define our goal and
to get a better aim at the targets. We need to go through this step
in order to establish a common language, agree on the common goals,
to explore different methodologies and build up the systematics for a
more fruitful expedition in this wild, exciting yet equally perilous
terrain.

In this process the reader will see how this new (hydrodynamics)
viewpoint and  (bottom-up) approach to quantum gravity would bear on
this issue. In fact, after
analyzing the primary issues involved, we confess we failed to see
why or how this issue of decoherence (in the sense of loss of quantum
coherence) in quantum gravity could be of such importance towards the
emergence of low energy spacetime physics.  Decoherence is a central
issue in the quantum to classical transition. The emergence of
spacetime with a manifold structure is, in this new view, more an
issue of micro to macro transformation. With dark energy hovering
overhead we may one day even see a paradigm shift that today's large
scale spacetime structure is fundamentally quantum.

Finally, we want to emphasize that this article is more in the nature
of comments than expositions, with the purpose of discussing points
rather than covering an area, much less presenting or dwelling on a
volume of systematic work. Companion papers are in progress
addressing related issues such as intrinsic decoherence
\cite{IntDecAH}, gravitational decoherence in astrophysics
\cite{GraDecAP} and in black holes \cite{GraDecBH}.

\subsection{Organization of this paper}

We begin in Section 2 with a discussion of the meaning of
decoherence in QG, and lay out a set of basic issues which one needs
to address for the consideration of this problem. We caution that
many common conceptions may not be valid or  relevant here, and that
as a result the discussion of issues in low energy physics being
affected by Planckian physics is not straightforward.
 In analyzing the aspects possible new features of low energy physics
 stemming from quantum gravity novelties, we distinguish
between two types of issues. First, there are issues originating
from general mathematical or physical reasoning and their underlying
assumptions.  The other type of issues is specific to the particular
quantum gravity approaches taken. To address them one needs to place
them in the larger framework which these approaches adopt
\footnote{This is just a warning that debates on these issues may
not be so fruitful without considering the viability of their
underlying framework. For lack of space we will not probe into the
theoretical schemes at the base, but only bring out some issues
common to them all.}. In Sec. 3, we comment on these issues in
relation to the long standing research program of Mavromatos et al
\cite{Mavr} and in Sec. 4 we focus on Garay's proposal on thermal
spacetime foam \cite{Garay1}.  Through these examples, we illustrate
the cris-cross of issues involved, in quantum field theory and
statistical mechanics, such as the assumptions behind the structure
of spacetime foams, low energy effective theory, origin and nature
of noise, the Brownian motion analog and the validity of the Markov
approximation, etc. In Sec. 5 we continue our discussions into two
more finer issues. We then summarize  the points made in this paper
into three key concepts, i.e., nonunitarity, nonlocality and
stochasticity and give a discussion of them in a broader scope.


\section{Decoherence in Quantum Gravity: Meanings and Issues}

Before addressing the issues involved, let us begin by giving some
careful deliberation on the meaning of decoherence in quantum
gravity.

\subsection{Three different Meanings}

If we agree that quantum gravity is a theory for the microscopic
structure of spacetime then the microscopic features are what we need
to deal with, more so than the quantum features. Using ordinary
physical objects as example, given a solid the first quest is to
uncover its atomic constituents, the second quest is to find out the
quantum nature of solids
which would lead us to phonons and other collective excitations
rather than atoms. Of course there are interesting quantum features
associated with the atom, but this is a quest different from the
first two. The mixing up of these two concepts is unfortunate but
perhaps understandable because we usually view any microscopic world
as described by quantum physics
 \footnote{Maybe so,  until recently:  Macroscopic objects can also show
quantum features as witnessed by experiments defining the new
emergent field of macroscopic quantum phenomena (MQP) \cite{Arndt}.
However, it is usually believed that macroscopic objects decohere
more readily than microscopic objects and they can be described
effectively through their collective variables as a classical object
\cite{CHY}.} If the quest is of the first kind  then there is
virtually no issue of decoherence in quantum gravity. For a long
time studies in quantum gravity is of the nature of the second kind,
i.e., finding ways to quantize the variables in general relativity.
This involves the assumption that GR arises as a Hamiltonian coarse
graining (essentially via a Born-Oppenheimer approximation as is
mentioned later) from the underlying theory and that the metric
corresponds to its true degrees of freedom. In effect, the
theoretical preconception of how GR emerges as a theory, defines the
strategy one will follow for its quantization \footnote{Quantization
assumes a function inverse to that of coarse-graining. A wrong
coarse-graining hypothesis may lead one completely astray. For
example, if the coarse-graining is hydrodynamic, a Hamiltonian
quantization would be as meaningless as quantizing the Navier-Stokes
(or the Euler) equations and arguing that the quanta of the mass
density are the atoms.}  The quantum theory thus obtained is assumed
to be a theory for the microscopic structures of spacetime. It seems
to us that this requires making some additional strong assumptions.
If one {\em assumes} that the key feature of gravity is geometry at
all energy and scale ranges, then the canonical quantization
procedure does make sense, even though it is not the only possible
strategy. A common assumption is the postulate of the existence of a
minimal quantum area, for example, is similar to specifying that a
solid of the size of an atom will give us the microscopic structure.
But this postulate serves to demarcate the domain of validity of a
solid's quantized modes of vibration, but says very little about the
atom from this theory of phonons, let alone its internal structure
or its quantum features. \footnote{For the same reason we don't
believe that the quasi-normal modes of a black hole which describe
its classical vibrations can be simply extrapolated to reveal the
microscopic structure of a quantum black hole or the `atoms' of
spacetime, as some earlier claims seem to suggest.}

\subsubsection{Quantum-Classical transition versus micro-Macro
manifestation}

To give the best reading of what practitioners could have meant by
decoherence in quantum gravity, and to make it more interesting and
meaningful we should perhaps take on the third quest: i.e., that it
is about the decoherence of the quantum features of the microscopic
constituents of spacetime, like the quantum features of the atom
after we make the discovery that solids are made of atoms.  Note that
our best reading also catapults us to a futuristic world where we
already know what the atoms of spacetimes are and we want to know how
the classical spacetime described by general relativity theory comes
into being. Some optimists would consider that future is now, the
atoms of spacetime are strings or loops or spin-nets. Fine. Let's
examine what issues we need to deal with in such a picture, and then
examine how the practitioners really think about them and what they
actually do.

To give a hint that this may not be the right question or the most
interesting question to ask,  let us return to the atoms in a solid
for a moment. This question pertains to how the classical features
of a solid could arise from the quantum nature of atoms. We know
there are more interesting questions in condensed matter physics
than this one: How the solid appears as we put the atoms together
depends on atomic bonds and that in turn depends on quantum dynamics
between atoms. One can also ask how the symmetries of crystals
emerge (be mindful that the art of crystallography predated quantum
physics). Or, how the thermodynamic laws in the macroscopic world
emerge.  Many classical features of solids can indeed be traced to
the quantum interaction amongst atoms, but they are not about
decoherence or recoherence. How does the classical nature of a solid
arise from the decoherence of quantum atoms is not really a burning
issue in solid state or atomic physics.

In gravitational physics, assuming strings  play the role of the atom
of spacetime, we see that the more urgent issues are to get the low
energy spectrum (note that the particles made from strings are
quantum objects), and how spacetime emerges \footnote{It may not be
too unfair to say that at this stage, even getting classical GR or
low energy particle spectrum out of strings or loops is a big issue.
In the landscapes scenario it seems to us that these real physical
issues are relegated to metaphysics -- \underline{how} do metastable
states come about, and \underline{why} we are in this particular
one?}. The focus is not so much on how the classical nature of
spacetime emerges from the decoherence of strings.  This would be a
`letdown' of sorts to those who believe they already have the
microscopic constituents at hand, because decoherence is about
degradation of information, and usually in an indiscriminate way, by
the action of an environment. The heavily laden statistical
procedures which underlie this process known as coarse-graining
almost does violence to the more important tasks we face, that of
constructing the macroscopic world from the precise information about
the constituents\footnote{One analogy is to take pride in discovering
the Maxwell-Boltzmann distribution at the high temperature limit of a
quantum gas, while ignoring the many interesting physics associated
with a bose or fermi gas.} In this sense, the discussion of gravity
induced decoherence at low energies is an indirect way to make
contact with possibly observable effects in the low energy world.


\subsubsection{Quantum vs Statistical}

Let us now focus on what the practitioners addressing this issue
really are referring to. Maybe they want to uncover the structure and
nature of spacetime at a scale smaller than the Planck scale (e.g.,
Garay's thermal spacetime foam \cite{Garay1}), or use a presumed
structure to deduce some hidden and yet important relations in
physics (such as the relation between clocks, computers and black
holes as in Ng's picture \cite{Ng}). Maybe they want to understand
the origin of fundamental relations in quantum mechanics, such as the
uncertainty principle (e.g., \cite{Milburn}), thinking that gravity
may play a hidden yet significant role \cite{Penrose}, even at
today's low energy. Maybe they think that certain symmetries
cherished at today's low energy can be broken by some quantum gravity
process \cite{Mavr}. An examination of the targets of these
investigations is useful perhaps not so much for the purpose of
finding out what decoherence can bring forth as how the issues can
play out in the quantum $\bigotimes$ gravity context.

Before we address these issues and try to extract their underlying
meaning, for concreteness, we need some entity or notion of an entity
to focus our thoughts.  In the realm of quantum gravity the word
spacetime foam or quantum foam are used often to connotate that phase
of spacetime at a higher energy than the Planck energy. Spacetime
foam as conjured by Wheeler refers to the primordial state of
spacetime made up of foamy structures of multiply-connected
topologies. There is in principle nothing quantum about it even
though in the pre-Planckian epoch most people would allow for quantum
fluctuations of spacetime to contribute to these foamy structures. We
make the distinction here between these two entities, calling the
former spacetime foam and the latter quantum foam. The reason is that
there are subtle and physical differences between statistical and
quantum entities.

Consider first  the statistical mechanics of classical foam-like
structures \cite{Hamber}.  In the statistical foam we have a {\em
single} `spacetime' manifold (or spatial three-surface) which at the
Planck scale is manifests a `foam-like' multiply-connected topology.
The statistical aspect rests on  how this topology averages at the
sub-Planckian scale and what are the effective variables that
describe this average. In the quantum foam, we have many different
`virtual' spacetimes, each with a different Planck-scale structure
and the issue is to determine the effective dynamics arising from
the contributions of all histories (in a path integral
representation). (For an interesting recent work on observables in
effective gravity, not necessarily related to spacetime foams, see
\cite{GidMarHar}).


What would decoherence in quantum gravity as embodied in
Planck-scale foam entail? For quantum foams this would pertain to
the dominance of the classical configuration. Some might argue that
one can transform a statistical problem to a quantum problem by
invoking the equivalence between a generating functional and the
partition function. But this invokes an Euclidean path integral
formulation or presumes a canonical ensemble, which in turn assumes
an equilibrium condition, as in thermal field theory. Also the Wick
rotation is not unique without a background notion of time--even in
special relativity one needs the full Poincar\'e group to make it
unique. The quantum fluctuation formulation usually presumes the
existence of a background spacetime, which has the smooth manifold
structure. So one cannot quite talk about the geometry of
trans-Planckian spacetime without a low energy background.

One useful way to connect to low energy physics is by way of
effective field theory, from which another range of problems enters.
We will have something to say in the next subsection.

Just these simple descriptions above which are familiar to most
readers bring forth a number of issues: 1) can one replace
statistical by quantum formulation or vice versa in these
considerations? If so argued, what are the conditions which validate
it?   2) How does coarse-graining of micro structures determine the
salient features of a large scale structure? 3) What gives rise to
stable low energy configurations? 4) Does decoherence at high energy
alter the symmetry at low energies?

\subsection{Issues}

\subsubsection{Decoherence in quantum gravity: an open system perspective}

How does one inject decoherence into this picture of quantum gravity?
The simplest way is to apply the popular environment- induced
decoherence scheme to spacetime foams. The first question one needs
to ask is, ``What could constitute the system? What the
environment?" Usually because of the `inertia' \cite{GelHar2} of the
gravity sector (due to the discrepancy between the Planck energy and
our ordinary energy -- see below)  one would be tempted to say that
the gravity sector is the system and the matter sector its
environment. If one wants to use this scheme, it should be done
without asking for the aid of a manifold structured spacetime,
because otherwise one is not addressing decoherence in quantum
gravity (in the third sense above) but decoherence due to classical
gravity below the Planck energy.

But what could the environment be in an earlier epoch, at the level
of the substructure? What are the criteria which separate a
subsystem from the others which can evolve into the macroscopic
spacetime we are familiar with? Decoherence due to Planck scale
(top--down) effects {\em that cannot be identified from the theory
at low energy} involves by necessity assumptions about Planck scale
physics.
To our knowledge these questions have not been yet been fully or
properly addressed.
\footnote{There are proposals of decoherence due to Planck scale
effects based on specific limits of `measurability' posed by gravity.
Many arguments start with Wigner's analysis of the spacetime
uncertainty relation, (e.g. \cite{Ng, Gambini}).  However,
fundamental uncertainties definitely do not imply decoherence by
themselves: they involve additional assumptions in order to do so.
For example, Milburn \cite{Milburn} models these uncertainties by a
stochastic process. In effect, these works treat the spacetime foam
as something that has properties similar to a classical source of
noise. Quantum coherence of the foam is ignored, as well as
non-Markovian effects. We will address the insufficiency of Markovian
assumptions in the last section and discuss some examples of this
so-called intrinsic or fundamental decoherence in an accompanying
paper \cite{IntDecAH}.}

\subsubsection{Spacetime structure at high energy from a low energy view}

The usual assumption is that somehow spacetime structure and quantum
fields emerge as an approximation at energies lower than the Planck
energy. The first question is in what sense these properties emerge,
or, what class of coarse-graining will likely give rise to these
properties, and what are {\em their} attributes?

At least three possibilities have been discussed in the literature:

{\it A. Born-Oppenheimer (BO) Approximation} (sometimes called
"Hamiltonian coarse-graining") is what we have referred to above: If
one looks at the `phase space' of the full theory at a coarser
resolution, those degrees of freedom with a bigger `inertia' (in the
sense of  Gell-Mann and Hartle \cite{GelHar2}), would appear to
behave classically--see also \cite{Omn}. The gravitational sector
being weighted by the Planck energy over the matter sector is what
justifies the introduction of a WKB time in quantum cosmology and
how it produces the limit of quantum field theory in curved
spacetime \cite{HarHaw83}. This is a kinematical rather than a
dynamical explanation for the transition from quantum to classical.
According to this perspective there can be no phenomena at low
energy due to gravity but by that described by the theory of general
relativity, because classical gravity dominates (is weighted
favorably in the GH sense, under the BO approximation) below the
Planck scale.

{\it B. Correlation hierarchy}  -- When there is a distinct
discrepancy (in time, length, mass) between two sectors one can use
the open system paradigm to describe the dynamics of the subsystems
of interest.  A more difficult situation is when no such discrepancy
exists as in the case of a molecular gas, where each gas molecule is
autonomous. Nonetheless,  Boltzmann taught us how to order the
information in the system,  in terms of one particle distribution
function, two particle correlation function, etc,  which form the
correlation (BBGKY) hierarchy. He also taught us how to introduce a
coarse-graining in accordance to the precision we can access the
information in the system, where the Boltzmann entropy acts as a
measure of the degree of ignorance of such. This might be a more
suitable way to view the pre-Planckian dynamics of the
sub-constituents, analogous to molecular dynamics before
hydrodynamics takes shape. There again, quantum decoherence is not as
important or urgent an issue as studying how the hydrodynamic limit
comes about, or from that state deduce the molecular properties.

{\it C. Effective theory and scaling of coupling constants} This
familiar concept is well adopted in particle theory to address the
hierarchy problem. The idea of scaling which works very well for the
study of critical phenomena also added much richness to this
paradigm. It allows one to talk about low energy phenomenology,
without paying too much attention to the underlying more fundamental
theory. The important parameters or scales are the thresholds where
the low energy effective theories break down, in ranges where new
particles and interactions become important.

A natural question to ask is ``What could have survived from the high
energy sectors?" Calzetta and Hu \cite{eftCH} answered this question
in terms of noises coming from the coarse-grained higher energy
sectors (which could be colored and multiplicative depending on the
interactions prevailing), but they fall off exponentially fast below
threshhold. One could try to decipher the degraded information from
the nature of noise, but overall this result affirms the philosophy
of effective field theory,  that it is not easy for high energy
processes to cause appreciable effects at low energies. This issues
about noise from effective theory will arise in Sec. 4.

\subsubsection{Breaking of symmetry at low energy due to decoherence
at high energy}

One hope some practitioners place on decoherence is that open system
dynamics are non-unitary: if such dynamics can be obtained from
quantum gravity, the resultant low energy dynamics would also be
nonunitary, thus providing a convenient way to break important low
energy symmetries, like CPT. This argument involves the physics in
two separate scales, a familiar low energy one endowed with a
symmetry and a causal spacetime structure, and another unknown one
at higher than Planck energy. This cannot be easily justified,
irrespective of the definition of decoherence one employs. For
example, in the usual environment-induced decoherence framework, the
open system may arises from the effects the ordinary gravity sector
at today's energy. However, in schemes such as the above, the
dynamics becomes non-unitary because of the influence of an
`environment' that involves trans-Planckian constituents engaging in
(unknown to us) quantum gravitational processes. There is an obvious
mismatch here, which we believe to be unphysical for the following
reasons:

Looking at the problem in the high energy realm,  any entity of
quantum gravity vein such as spacetime foam is at the Planck
scale--supposedly before spacetime with a Lorentz structure emerges.
One needs a strong case to show that these high energy effects can
escape the coarse-graining and scaling which subsumed their effect
to that of the average, as the large scale manifold structure of
spacetime emerges. In our opinion, the most reasonable assumption is
that their average behavior is contained in the theories that
survive to the low energy limit, in particular GR--it appears very
counterintuitive to assume otherwise. If this is the case, the
non-unitary corrections cannot be of order $G$--effects to this
order are contained in the usual gravitational dynamics--but much
much smaller.

Looking at the problem from the low energy realm, the coupling of
matter to gravity is through the stress-energy tensor $T_{mn}$. If
the Lagrangian of matter is invariant under a symmetry, this would
be reflected in the $T_{mn}$. There is no way to get a  symmetry
violation, unless it is already there at the low energy theory.
Moreover, if there is non-unitarity due directly to Planckian
effects, it would not affect specifically one type of symmetry, but
all of them: hence one should expect besides CPT violation, a
violation of the spin-statistics relation, of the causality
properties of quantum fields, and perhaps of less universal
symmetries like baryon and lepton number.
For these reasons, we believe that the study of gravitational
decoherence at low energies should primarily focus on the physics of
GR: Penrose-type reduction, collapse models, gravitons as environment
etc. Otherwise, one would have to answer the very difficult question
``Through what mechanism would Planck scale effects {\em dominate}
over ordinary gravity effects at this low energy?\footnote{The issue
of dominating over low energy theories is important. For example, the
conjectured `large' extra dimensions could presumably be sources of
symmetry breaking, but their contribution to  gravitational dynamics
responsible for decoherence are expected to be small compared to
ordinary GR.} " A postulate of exotic open system dynamics can only
be justified if it is tied to a concrete theory of quantum gravity.

\section{Quantum Gravity Decoherence and CPT violation}

Continuing on the last point, we comment in this section on a
related proposal \cite{Mavr}, which evokes spacetime foams, in order
to provide rationales for  results in high-energy physics
phenomenology (`high energy' here does not refer to Planck scale).
The suggestion is that the presumed non-unitarity from Planck-scale
processes may appear in particle physics experiments, e.g. neutrino
or neutral meson oscillations.

Two classes of models have mainly been employed in this regard. The
first is a Markovian master equation for the distinguished degrees
of freedom, flavor for neutrinos--for three generations the Lindblad
operators may correspond to generators of the $SU(3)$ group
\cite{Mavr2}. The second approach involves the study of neutrino
oscillations in a stochastic medium (Mikheev-Smirnov-Wolfenstein
effect \cite{MSW}), which they identify with fluctuations from the
spacetime foam.

While the results of this analysis can fit the experimental data, we
would like to make some points concerning the relation any such
presumed decoherence effects to gravity. First, even if it turns out
that there is a decoherence effect in neutrinos, it would be
premature at this stage to attribute it solely or even dominantly to
gravity. Any kind of environment due to higher energy processes that
may be involved in the weak interactions could play the role of a
decohering agent. Second,  gravity (as described in general
relativity) couples universally to all types of matter and to all
degrees of freedom that contribute to the stress-energy tensor:
hence, the isolation of specific degrees of freedom cannot be
justified from first principles. In the case of neutrinos, the
universality of gravity would suggest a significant coupling between
the flavor and the translational degrees of freedom. If there are
specific processes that distinguish the flavor degrees of freedom,
present knowledge suggests that their origin lies in weak
interaction physics, not in gravity. Finally, there is also the
issue of the Markov assumption, or of the modeling of the stochastic
background: as we will argue in Sec. 5 the possible behaviors of the
Planck scale foam are not exhausted in the Markovian regime and
there is no guarantee that they can be modeled adequately by a
stochastic process.

\subsection{CPT violation}

Another argument in \cite{Mavr} is that the non-unitarity effects
arising from quantum gravity may lead to CPT violation in high
energy physics experiments. The argument is based on the fact that
the CPT theorem requires Lorentz covariance and unitarity. If the
effective low energy dynamics is non-unitary or non-Lorentz
covariant then one of the basic conditions for the CPT theorem is
violated: CPT violation is then plausible.

We urge the exercise of caution on this point. The fact that the
effective dynamics is non-unitary is not sufficient to guarantee CPT
violation. To see this, one may consider a low energy theory (say
QED), in which the soft photon modes have been traced out. The
effective dynamics of the theory is non-unitary. However, there is
no CPT violation in the effective description: photons have no
anti-particles; left-handed and right-handed photons are treated
symmetrically in QED. The channel of interaction of the system to
the environment involves no CPT violation. The same would hold for
any field coupled to the gravity, if one traces out the effect of
gravitons: gravitons have the same CPT properties with photons, and
the effective dynamics they generate when treated as an environment,
fully respect CPT.

In other words, for the effective non-unitary dynamics of a low
energy theory to violate CPT, it is necessary that the channels of
interaction with the environment (which are represented by the
Lindblad generators in the Markovian regime) are themselves not CPT
invariant. This can happen only if the total Hamiltonian of system
and environment involves CPT violating terms. Hence, the possibility
that the effective dynamics at low energy is non-unitary does not
suffice to establish  CPT violation. One has to identify specific
physical processes (i.e. terms in the total Hamiltonian) that do so.
Moreover, even if there is CPT violation near the Planck scale,
there is no guarantee that its effects will be felt at low energy.
In our opinion, the most natural assumption is that such terms would
take the form of small corrections to the combined predictions of
general relativity and standard model. The (by far) dominant
contribution to gravitational decoherence at low energies would then
come from the non-CPT violating effects: the graviton environment or
 other suggested low-energy gravity mechanisms (e.g. Penrose-type).

\section{Quantum Foam and Emergence of Classical Spacetime}

We mentioned earlier that the emergence of spacetime and low energy
physics is often viewed as a sequence of effective field theories
each of them valid at a different scale. A necessary assumption in
this approach is that somehow after the Planck scale a full-fledged
spacetime structure with its basic properties has emerged -- it is
technically impossible to work in absence of this. The basic
variables in an effective field theory are not required to be and
often not the same as the fundamental variables of full quantum
gravity. In condensed matter or nuclear physics, the effective
degrees of freedom may be collective or hydrodynamic.  We have to
work much harder to decipher how attributes of a Planck scale entity
such as the spacetime foam could enter into the low energy physics
we are familiar with. We mentioned before the two features which
could carry these remnant information, such as nonlocal noise
(through the correlations of the collective variables) and
topological structures (likely convoluted by intervening processes).

In the present context, the question is whether the effective theory
for physics after the emergence of spacetime may cause observable
effects (in particular, decoherence) at low energies that differ
from the ones obtained from our known low energy theories (i.e.
standard model + GR). {\em A priori}, this is not very plausible.
There are energy thresholds, and effects appearing {\em only} in
energies above a specific threshold are strongly suppressed in
energies below the threshold.

In Refs. \cite{Garay1, Garay2}, Garay explores a different
alternative, namely that the dynamics of the effective degrees of
freedom after spacetime has emerged are non-local. This involves a
rather non-trivial assumption, namely that local dynamics is a
distinct property (and it arises at a different scale) from the
spacetime structure that is necessary in order to phrase a quantum
field theory meaningfully. While this separation is mathematically
sound, its physical meaning is not straightforward: at least in GR
it is difficult to separate between the local structure of spacetime
and the locality of dynamics. The locality of the action in general
relativity is a necessary feature that allows one to connect the
mathematical objects ``metric" and ``spacetime point" with physical
geometry; for example, locality is necessary in order for
free-falling particles to move in geodesics. But even outside the
context of general relativity, continuity of the effective variables
is closely tied to locality of dynamics. This is the case in
ordinary hydrodynamics: there is no way to separate the regime, in
which the continuum approximation holds, from the regime of local
dynamics; one assumption does not make sense without the other.

In Ref. \cite{Garay1} a separation of three scales is assumed: the
Planck scale $l_*$, the scale of the gravitational fluctuations $r$
and the scale of low energy physics $l$. Let $\phi$ denote the
variables of the effective theory, and $h_i[\phi](t)$ denote a basis
of local interactions at a spacetime point $(x,t)$. Then one
introduces into the action a term of the form $I = \sum_N I_N$,
where $I_N$ is an $N$-local interaction term
\begin{eqnarray}
I_N = \frac{1}{N!} \int dt_1 \ldots dt_n c^{i_1 \ldots i_N} (t_1,
\ldots t_n) h_{i_1}(t_1) \ldots h_{i_N}(t_N).
\end{eqnarray}
The functions $c^{i_1 \ldots i_N} (t_1, \ldots t_n)$ should not
depend on the location of the gravitational fluctuations, hence they
should be functions the relative location of the interactions.
Moreover, they should vanish if the relative distance is
substantially larger than the scale $r$. It is reasonable to assume
that fluctuations involving a large number of spacetime points are
suppressed; hence  at first (weak-coupling) approximation one may
keep only the bilocal terms.

Assuming that the dynamics is described by the Euclidean path
integral, and keeping only the bilocal terms we obtain an expression
$\int D \phi \exp[- I_0 - \int dt dt' c^{i j}(t - t') h_i(t)
h_j(t')$, where $I_0$ is the action part that describes  the
self-dynamics of the effective degrees of freedom. Using a standard
manipulation of the quadratic interaction term, together with a Wick
rotation back to Lorentzian spacetime, one obtains the following
expression for the path-integral
\begin{eqnarray}
\int D \alpha P(\alpha) \int D\phi e^{iS_0 + i \int dt \alpha^i(t)
h_i(t)}, \label{pathint}
\end{eqnarray}
where $\alpha$ is an auxiliary variable and $P(\alpha)$ is a Gaussian
probability measure
\begin{eqnarray}
P(\alpha) = \exp[ - \int dt \int dt' \gamma_{ij}(t-t') \alpha^i(t)
\alpha^j(t')],
\end{eqnarray}
where $\gamma_{ij}$ is the operator inverse of the kernel $c_{ij}$.
Eq. (\ref{pathint}) essentially describes unitary propagation under a
stochastic external force $\alpha^i(t)$, which fluctuates according
to a classical stochastic measure $P(\alpha)$.

Eq. (\ref{pathint}) is analogous to similar expressions appearing in
the study of quantum Brownian motion QBM (even though the exact
interpretation of this object depends on the choice of boundary
conditions for $\phi$). In fact, as Garay argues in \cite{Garay2},
one expects that the most general possible bilocal dynamics can be
expressed in terms of an influence functional
\begin{eqnarray}
W[\phi, \phi'; t]  = \exp\left\{ -\frac{1}{2} \int_0^t ds \int_0^s
ds' \left(h_i[\phi(s)] - h_i[\phi'(s)] \right) \left( v^{ij}(s-s')
h_j[\phi(s')] - v^{ij}(s-s')^* h_j[\phi'(s')] \right) \right\},
\label{IF}
\end{eqnarray}
in terms of a complex-valued kernel $v^{ij}$.

Following (\ref{pathint}) one may derive a master equation assuming
an ensemble of unitarily evolving systems, each acted upon by an
external force, whose ensemble average is provided by $P(\alpha)$. At
lowest order in $r/l$, the result is
\begin{eqnarray}
\dot{\rho} = - i [H_0, \rho] - \int_0^{\infty} d \tau c^{ij}(\tau)
[h_i, [h_j, \rho]]. \label{maseq}
\end{eqnarray}
Note that this master equation does not contain any dissipation
term, which seems to be of a higher order in $r/l$. This expression
is then compared with one obtained from QBM in a thermal bath of
harmonic oscillators with a $h_i$-coupling to the effective
variables $\phi$ \cite{Garay1}. The conclusion is that the leading
behavior at the classical noise limit (when the commutators of the
bath variables can be ignored) is the same as in (\ref{maseq}), and
hence that it is meaningful to talk about the spacetime foam as a
thermal quantum bath.

We want to argue that while it is reasonable (in fact, natural) to
compare the phenomenology of non-local effective theories to quantum
Brownian motion, the conclusion that the assumed non-local
fluctuations of spacetime foam behave similarly to a thermal bath
involves specific modeling assumptions and cannot be considered as
definitive.

Concerning the first point, we agree that the quantum Brownian
motion provides the most natural framework for the discussion of
non-localities in quantum gravity. In fact, using the QBM language,
it is not necessary to postulate fundamental non-localities at
sub-Planckian scales, which may bring about issues of causality
violation extending to low energy physics. The reason is the
following.

The variables $\phi$ that appear in the effective field theory are
probably collective degrees of freedom arising out of the Planck
bath. The existence of non-localities in their dynamics is
essentially the statement that the foam bath exhibits intrinsic
correlations of characteristic scale $r$. We can compare it with the
analogous situation in ferromagnetism: the basic {\em local}
thermodynamic variable is the magnetization, however in some regime
(e.g. near the phase transition) the correlations of the system
become important and in some phenomena they contribute significantly.
For this regime, one can write an effective non-local Hamiltonian for
the magnetization, but one can equally well introduce new
phenomenological fields. Alternatively, one may say that the true
degrees of freedom are not the local fields, but some specific
combination of them (one would consider for example an analogy with
magnons).

In effect,  the non-local interactions are equivalent to local ones
that involve  additional phenomenological fields $A^i$. They should
correspond to a correlation length of order $r$, which is equivalent
to them being characterized by a mass $M$ of order $r^{-1}$.

For the total system of phenomenological fields $A^i, \phi_i$, it is
sufficient to postulate a local interaction term $\int dt A^i h_i$.
The effective interaction between the $\phi$ fields will then be
similar to $I_2$ considered in Garay's model -- the role of the
kernel $c^{ij}$ will be played by the two-point function of $A^i$ --
as long as the effective theory is viewed from scales larger that
$r$. However, in scales of order $r$ the dynamics of the fields $A^i$
will be significant, and a non-local action like $I_2$ may not be
adequate to describe physics at this scale. Still, as argued in
\cite{Garay2}, an expression like that of the influence functional
(\ref{IF}) will be relevant.

Since the fields $\phi$ involve degrees of freedom that are
accessible to low energies, and the dynamics of the $A^i$ fields are
frozen at low energies it is meaningful to trace out the
contribution of the latter. In effect, the problem reduces to that
of QBM with the environment consisting of relatively heavy fields.
Hence, QBM provides a more general paradigm for the treatment of
non-localities than the postulate of a specific non-local effective
interaction. The issue is then what is the physically relevant state
of the bath. If the $A_i$ fields are similar to fields of low energy
physics (e.g. scalar fields), and if we assume that they can be
described by a thermal state of temperature $T$, for $T << r^{-1}$,
an effective dynamics described by a version of $I_2$ is plausible,
while for $T \sim r^{-1}$ the dynamics will be of the more general
form (\ref{IF}).

However, we believe that the identification of spacetime foam with a
thermal bath involves specific modeling assumptions for the
environment, and it is not a necessary conclusion. The reason is the
following. In standard QBM, different assumptions about the
dynamical properties of the environment (in particular self-dynamics
and initial state) lead to very different behavior of the reduced
degrees of freedom. The analysis is often simplified by invoking the
Born approximation, i.e. assuming that the back-action of the system
to the bath is negligible. However, this approximation is not always
adequate: for a Planck scale bath, in particular, it involves very
strong assumptions about the physics of spacetime foam. In general,
the consideration of back-action leads to stronger correlations
between the system and the bath, which are not fully contained in a
master equation \footnote{As shown in \cite{PazZur}, the evolution
law of the reduced density matrix does not in general capture the
temporal correlations of the open system: a sufficient condition is
that the reduced dynamics is time-homogeneous Markovian--see also
\cite{CRV}.}. For some environments, these correlations are
preserved even in the semi-classical regime and they affect
significantly the evolution of specific observables. In such cases,
open system dynamics do not lead to decoherence.

In particular, a Markovian master equation that leads to decoherence
for the reduced degrees of freedom [say, of type (\ref{maseq})]
arises only under the assumption of very specific conditions for the
bath. A weak coupling assumption is necessary (which typically does
not hold in strongly correlated environments), but also an assumption
that the quantum correlations in the bath are suitably suppressed.
For example, in the special case of a thermal bath of harmonic
oscillators, it is only in the high-temperature regime and with an
Ohmic distribution of frequency modes that the resulting master
equation is Markovian. All other regimes are non-Markovian and
decoherence therein is generally less effective. For these reasons,
we believe that in an open system analysis of the Planck
fluctuations, the thermal behavior of the quantum foam arises only in
specific regimes for its internal dynamics.

Finally, we note that at the low energy limit the coupling of matter
to gravity is through the stress energy tensor. If we assume that
the phenomenological fields $\phi$ reduce to the ones of low energy
physics, one would expect that the coupling terms $h_i[\phi]$ should
reduce at low energy to components of the stress-energy tensor for
matter, irrespective of its specific form near the scale $r$. Here
the roles of the metric fluctuations and the matter fields are
reversed as compared to stochastic gravity \cite{stogra}, where the
matter fields are regarded as the environment of the metric
perturbations. In that case, the metric perturbations appear as the
argument of the influence functional whereas the kernels involve
two-point quantum correlation functions of the stress tensor
operator of the matter fields.

\section{Further issues and Key concepts}

We now conclude with a discussion of two more issues which could be
overlooked, or dealt with out of convenience than from principles.
They are the universal coupling of gravity and the Markovian
approximation. We end with a revisit of the key concepts discussed in
this paper, that is, non-unitarity, non-locality and stochasticity,
which we believe play a fundamental role in this class of problems.

\subsection{Two More Issues}

Two common assumptions  about the structure of the gravitational
degrees of freedom are made in many schemes of quantum gravity,
because they simplify the calculations enough to produce some
results. But we have not seen too much physical justification for
these assumptions.

 The first assumption concerns the nature of open-system dynamics
generated by gravitational degrees of freedom, namely that they are
Markovian. The second type of assumption involves the {\em a priori}
isolation of specific degrees of freedom that are decohered by
gravity and comes in conflict with the universality of the
gravitational coupling.

\subsubsection{Results under Markovian approximation are not generic}

To begin with, open system dynamics contains memories of its past
from the backreaction of its environment which has a different set
of time scales. Thus it is generically non-Markovian. There are
specific conditions related to the appearance of Markovian behavior,
e.g., the response time of the bath must be much shorter than the
dynamical time-scale of the system. In QBM, the response time is
related to the effective cut-off of the excited frequencies for the
vacuum, or to the inverse temperature of a thermal bath.

If we consider a graviton bath, there is no physical justification
to assume that it lies in a thermal state, because gravitons
thermalize very weakly, at least now. Since the coupling to gravity
is very weak, the excited frequencies are small, and by all
reasonable estimates the corresponding time-scale is very large in
comparison to the ones characterizing the motion of the particle--at
least as long as we do not consider planet-sized objects. Hence, the
separation of time scales goes the opposite direction of that in
ordinary Brownian motion.

If one tries to tie decoherence to spacetime foam, i.e. highly
non-linear Planck processes, the issue of light-cone fluctuations
comes into account. It is not clear how to interpret time there,
much less assume that the dynamics have specific properties with
respect to time. In fact, the notion of Markov behavior may not make
any sense at all in the Planck scale.

\subsubsection{Gravity coupling is different from how other fields are coupled}

According to  GR the gravitational coupling is universal: the
channel through which matter interacts with gravitons is always the
stress-energy tensor. There is no room for manoeuvre in that issue:
for any model to connect to reality it must take this fact into
account. One should therefore be sceptical about any ad-hoc
assumptions involving isolation of specific degrees of freedom,
specification of convenient open-system dynamics etc. As long as one
trusts the equivalence principle, universality is a fundamental
feature of gravity and this should be mirrored also in the treatment
of decoherence. (Even if the equivalence principle is violated, the
violations will only be small corrections to the dominant
contribution that is encoded in the stress-energy tensor.)

The specific coupling of gravity to matter through  the
stress-energy tensor $T_{mn}$ is quite different from that of other
fields. Even if the assumption of a Markov dynamics were acceptable,
the choice of the Lindblad operators cannot be arbitrary. They must
be obtained from the stress-energy tensor. Moreover, for extended
systems it is not necessary that the Lindblad operators commute with
energy (as it is sometimes assumed). This may be the case for single
isolated particles (for example, see \cite{Ana96}), but when one
considers many-particle systems or fields (especially massless ones)
there is no reason for this assumption to hold. (There is definitely
some energy loss due to the quadrupole moment).

\subsection{Key concepts: Nonunitarity, Nonlocality and Stochasticity}
\vskip .2cm

{\bf Non-unitarity in sub-Planckian vs trans-Plankian physics}

{\it 1. Decoherence in Quantum Gravity and Non-unitarity in
sub-Planckian physics}

We commented in Sec. 2.2.3 and Sec. 3 about the issue of using
nonunitarity in open systems (dissipative dynamics) as a source or
explanation for the violation of symmetries in physics at the
sub-Planckian scale. For completeness we add here an old
consideration of non-unitarity in quantum gravity stemming from
black hole physics.

{\it 2. Non-unitarity in Quantum Gravity from black hole evaporation
arguments}

A usual argument in support of decoherence at low energies is the
hypothesis that the dynamics of quantum gravity is non-unitary, which
is motivated by the consideration of the black hole evaporation
process. We recall here the argument: matter in a pure state
collapses and forms a black hole; the black hole slowly evaporates by
emitting thermal Hawking radiation; after total evaporation there
will be no horizon and the information of the initial state will have
been lost: a pure state will have become mixed. Hence, the dynamics
of quantum gravity must be non-unitary.

Setting aside the issue whether the non-unitarity of quantum gravity
is the only solution to the black-hole evaporation `paradox', we
need to point out that the same process also involves violation of
certain important conserved quantities of high-energy physics, like
baryon number. If the only motivation for assuming non-unitary
dynamics is the above, then one would have to expect that the
non-unitary quantum gravity processes would be manifested together
with baryon number violation. Otherwise, one would have to assume
distinct underlying physical mechanisms that cause these phenomena
in black hole evaporation. Hence, if decoherence effects at low
energy are attributed to the same cause as the ones related to black
hole evaporation, one should also expect a violation of baryon
number of the same order of magnitude.

Moreover, at low energies there are different (non-Planck scale)
processes that may be responsible for gravity-induced decoherence.
The graviton vacuum acts as a universal bath for all systems and to
some degree this may also be the cause for effective open system
dynamics for matter. There is also the suggested Penrose mechanism
\cite{Penrose}, whose effects can be identified even at the level of
Newtonian gravity. It seems to us that these mechanisms will be more
efficient in causing decoherence (if they do that) than presumed
Planck scale mechanisms, whose effect is more likely to be
suppressed
at low energy.\\

\noindent {\bf Nonlocality in open systems vs in quantum mechanics
and general relativity}

This is a complex and difficult issue. We can only make a few
comments here on different types of nonlocality.  Nonlocality in
quantum mechanics of the EPR like is probably the best known. It is
tied to measurements being of a local nature and measurement
instruments of a classical nature. This is taken up in earnest in
quantum information and communications. Then there is nonlocality in
time in open systems, i.e., memories in the form of nonlocal
dissipation and colored noises.  Nonlocality in space shows up in a
number of current theories about fundamental particles, strings and
fields. (For a discussion of nonlocality in gravity and string
theory, and black holes see, e.g., \cite{nonlocGra} and references
therein.)

It is sometimes argued that the `non-locality' may arise from the
presence of horizons in the spacetime foam, which manifest a
`non-local' behavior. Indeed, here is a sense, in which even ordinary
GR is `non-local' because of general covariance
\footnote{``Non-local" is perhaps not the most accurate denotation --
``sensitive to global structure" is more appropriate. To make a local
measurement of the geometry, one needs to fix a reference frame. To
do so, and interpret these results in terms of distant frames of
reference it is necessary to know the geometry outside the location
of the measurement: this is important especially when the asymptotic
flatness approximation cannot be employed. This is very different
from theories on a background spacetime, in which the set-up and
interpretation of an experiment needs only involve local knowledge.}.
We should note, however, that this `non-locality', like that of QM,
fully respects causality. Effective theories like the ones suggested
in \cite{Garay1} most likely do not: they seem to involve
`instantaneous' transmission of information at a scale where the
spacetime manifold has already emerged.

One may argue that any effective field theory is non-local.
Non-localities at some length scale $r$ larger than Planck length
may survive at low energies. However, the non-locality assumption is
essentially equivalent to the introduction of an additional
(universal) field of mass $r^{-1}$ that interacts locally with the
remaining degrees of freedom. One then traces out the contribution
of this field: we essentially have a QBM-type of situation with a
specific environment.  Attributing this effect to decoherence due to
gravity is a matter of interpretation:  Any (unobservable) field
could play the same role. Viewing effective field theory from an
open system viewpoint \cite{eftCH} one can easily see that the lower
hierarchy will have nonlocal dissipation on in its dynamics driven
by colored noises due to its interaction with fields higher in the
hierarchy acting as an environment to the low energy sector.
\\

\noindent {\bf Stochasticity: different appearance at different
levels of structure}

Any discussion of spacetime foam relates inevitably to the notion of
spacetime fluctuations. There are many arguments, ranging from simple
dimensional analysis to more complex ones that involve specific
approaches to the quantization of gravity, that the description in
terms of the spacetime continuum breaks down at the Planck scale and
that the concepts we employ in low energy physics arise only as
approximations. In that sense, the underlying quantum theory will be
manifested in the form of strong fluctuations for the effective
quantities.

This is how the notion of randomness or stochasticity is often
invoked in quantum gravity, notwithstanding our ignorance of what
this theory is. Wheeler's spacetime foam started this. Random
geometry in the 80's related to conformal field theory has made great
advances. If we stick to the definition that quantum gravity is about
theories for the microscopic structure of spacetime, then it is
plausible to think about a stochastic stage intervening between the
micro and the macro, much as hydrodynamic fluctuations from molecular
dynamics. In addition to its intrinsic features, one needs to factor
in the level of structure one is focussing on and the scope and
precision of observation into the system. Let us examine what
stochasticity entails.

Fluctuations could carry different meanings in different contexts:
deviations from a deterministic evolution, statistical fluctuations
in a many-body system, the limits to the definability of specific
phenomenological (or emergent) quantities. These notions refer
largely to the intrinsic behavior of the system and each of them
applies to a different physical circumstance.  There is, however, a
different use of the word "fluctuation" as it appears, for example,
in the context of measurement theory: it refers to all possible
external factors that can influence the outcome of an experiment or
simply the statistical spread in different trials or samplings. This
notion of fluctuation is {\em extrinsic} to the system under study,
and is usually represented by an external noise or random statistical
distribution. It is a usual practice in applied probability theory to
simulate the effects of external noise with stochastic processes. One
may employ general theorems or criteria to select a suitable process
that will give good agreement with the observed statistics.

At the Planck scale, many or all of these factors could enter into
one's consideration, intrinsic --  pertaining to the interaction of
the constituents,  and extrinsic -- pertaining to how the
information is extracted by the observer, as in our present
consideration,  at very low energies or later times. For example,
spacetime foam is intrinsically a geometric structure with
non-trivial topology. One needs to take that into account {\it ab
initio}. One could posit that at large scales this structure may
{\em appear} to be smooth and the emergent aggregate has a trivial
topology which joins with the manifold structure of classical
gravity nicely. Here stochasticity of an extrinsic nature enters, in
how a coarse-grained limit would appear to the low energy observer.
This latter set of issues are more subtle and just as important as
the intrinsic ones, because it enters in all considerations of
micro-macro manifestations. (For a discussion of how different
coarse-grainings bring forth different structures of varying
stability and robustness, see \cite{Timeasy}.)

One may be tempted to apply these same statistical notion of
fluctuations to spacetime foam.  Spacetime foam has a structure of
its own, independent of how we low energy creatures try to describe
it. Its structure and dynamics depend on the physics at the Planck
scale, period. We invent easier ways such as fluctuation theory to
try to capture its essence, but without due consideration of its
microscopic attributes, this could be a self-fulfilling prophecy
because the results are either {\em trivial} (often invoking the
central limit theorem) such as a thermal bath for something
stochastic, or {\em circular}, because the result is ingrained in
the assumptions.  When we introduce probability arguments such as
assigning a certain type of noise to capture its gross feature, we
are introducing extraneous information into its description which
not only could be totally off the mark but defeats the purpose of
our investigation (An example is using white noise to describe
strongly correlated systems -- it is doomed to fail from the
beginning because it is contrary to its spirit.)

We have warned against randomly invoking statistical fluctuations.
Now we add  the quantum aspect. Intrinsic fluctuations at the Planck
scale are believed to be quantum mechanical in origin. Thus the
terminology `quantum foam'. However, quantum effects cannot in
general be described in terms of stochastic processes.  Coherence
and nonlocality (e.g., Bell's theorem) are inherent quantum
properties which are lost in a stochastic description. Quantum
theory involves `interference' effects which do not allow the
definition of a stochastic measure--for different aspects of this
see, for example \cite{GelHar2,Omn, var, Ana01}.

Many proposed schemes involving fluctuations of spacetime
conceptually invoke a hybrid picture. That is, while insisting that
it is a quantum entity (otherwise it is not about quantum gravity)
one treats the foam-like structures as classical stochastic sources.
This is in fact the hidden motif in many quantum gravity decoherence
schemes \footnote{There are admittedly special regimes in physical
system where the distinctively quantum features such as coherence
are safely suppressed: the high temperature ohmic bath regime of
quantum Brownian motion is an example. In such cases, one may employ
classical probability and model the fluctuations through classical
stochastic processes. However, one needs to justify this on a
case-by-case basis. It is by no means generic. For QBM in a
supra-Ohmic bath at low temperatures, quantum coherence persists
much longer, as does quantum entanglement between the system and the
bath \cite{HPZ}.}.

Likewise, we cannot invoke the existence of a separate classical
regime. The fluctuations are effectively part of the classical world
which describes our low energy physics. A classicalized spacetime
foam is very efficient as a decohering agent; a fully quantum one may
be not. (This brings back our earlier discussion on the relation of
statistical versus quantum.) We see a circular argument here: this
classical world comes about from decoherence in quantum gravity but
the source of decoherence comes from statistical fluctuations of
classical geometries.

To summarize, there is no {\em a priori} argument why the spacetime
fluctuations should be modeled by a stochastic process, even though
in many situations stochasticity can be seen as a limiting sub-case
of quantum behavior. In a fundamentally quantum system (like the
spacetime foam) it cannot be assumed without a justification {\em in
terms of the physics of the system at the appropriate scale}, not by
the imposition of an external stochastic source.

To end, no discussion of quantum gravity can be complete without
bringing forth the issue of time. We postpone it to the end not
because it is not important. Time is implicitly assumed in any
description of dynamics, deterministic or stochastic, spacetime or
matter. It is because we don't have anything intelligent to say
beyond the familiar. We can only express our opinion: We would like
to see time and the causal properties it brings forth as emergent
from the interaction of microscopic constituents, together with
spacetime endowed with a manifold structure, but not as an extrinsic
element outside of the micro-system \footnote{An example is the time
scale or even the arrow of time in dynamics of collective variables,
such as hydrodynamics and thermodynamics. They are only indirectly
related to the time (without the arrow) in molecular dynamics. It is
enlightening to extract these macro features from the micro
structures in the various proposals (footnote 1).}.
\\

{\bf Acknowledgment}  BLH wishes to thank Professor Thomas Elze for
his invitation and hospitality to this interesting  DICE06
conference. We thank Dr. Albert Roura for helpful discussions on a
range of problems in
gravitational decoherence. This work is supported in part by NSF grant PHY-0601550. \\

\end{document}